\documentclass[12pt,thmsa]{article}

\usepackage{cite}

\begin{document}

\title{Calculation of bound states and resonances in perturbed Coulomb models}
\author{Francisco M. Fern\'{a}ndez \\
INIFTA\ (Conicet, UNLP), Divisi\'{o}n Qu\'{i}mica Te\'{o}rica,\\
Diag. 113 y 64 (S/N), Sucursal 4, Casilla de Correo 16,\\
1900 La Plata, Argentina \\
E--mail: fernande@quimica.unlp.edu.ar}

\maketitle

\begin{abstract}
We calculate accurate bound states and resonances of two interesting
perturbed Coulomb models by means of the Riccati-Pad\'{e} method. This
approach is based on a rational approximation to a modified logarithmic
derivative of the eigenfunction and produces sequences of roots of Hankel
determinants that converge towards the eigenvalues of the equation.
\end{abstract}

\section{Introduction \label{sec:intro}}

In a most interesting series of papers Killingeck et al \cite
{KGJ04,KGJ05,KGJ06} and Killingbeck \cite{K07,K07b} have shown that
perturbation theory and the Hill--series method are suitable tools for the
calculation of bound states and resonances of simple quantum--mechanical
models. In order to obtain the complex eigenvalues that correspond to
unstable states they resort to a complex parametrization of the methods that
they call ``complexification ''.

Another approach that proves useful for the accurate calculation of bound
states and resonances is the Riccati--Pad\'{e} method (RPM) based on a
rational approximation to a modified (or regularized) logarithmic derivative
of the eigenfunction \cite{FMT89a,FMT89b,F92,FG93,F95,F95b,F95c,F96,F96b,F97}%
. In this paper we apply the RPM to the interesting perturbed Coulomb
problems discussed recently by Killingeck \cite{K07b} with the purpose of
challenging the recently developed asymptotic iteration method \cite
{CHS03,F04,CHS05,CHS05b,F05,B05,B05b}.

In Section~\ref{sec:RPM} we outline the main features of the RPM. In Section
\ref{sec:boundstates} we discuss a perturbed Coulomb model with interesting
bound states. In Section~\ref{sec:resonances} we calculate the resonances
for a slightly modified model with continuum states. Finally, in Section~\ref
{sec:conclusions} we draw conclusions on the performance of the RPM.

\section{The Riccati--Pad\'{e} method (RPM)\label{sec:RPM}}

Suppose that we want to obtain sufficiently accurate solutions to the
eigenvalue equation
\begin{equation}
\psi ^{\prime \prime }(x)+Q(x)\psi (x)=0,\;Q(x)=E-V(x)
\label{eq:Schrodinger}
\end{equation}
where $Q(x)$ can be expanded as
\begin{equation}
Q(x)=\sum_{j=0}^{\infty }Q_{j-2}x^{\beta j-2}  \label{eq:Q_series}
\end{equation}
about $x=0$. We transform the linear differential equation (\ref
{eq:Schrodinger}) into a Riccati one for the modified logarithmic derivative
of the eigenfunction:
\begin{equation}
f(x)=\frac{s}{x}-\frac{\psi ^{\prime }(x)}{\psi (x)}  \label{eq:f(x)}
\end{equation}
On substituting (\ref{eq:f(x)}) into (\ref{eq:Schrodinger}) we obtain
\begin{equation}
f^{\prime }(x)+\frac{2s}{x}f(x)-f(x)^{2}-Q(x)-\frac{s(s-1)}{x^{2}}=0
\label{eq:Riccati}
\end{equation}
We choose $s(s-1)=-Q_{-2}$ in order to remove the pole at origin, and, as a
result, the expansion
\begin{equation}
f(x)=x^{\beta -1}\sum_{j=0}^{\infty }f_{j}x^{\beta j}  \label{eq:f_series}
\end{equation}
for the solution to the Riccati equation (\ref{eq:Riccati}) converges in a
neighbourhood of $x=0$. Notice that if we substitute the expansions (\ref
{eq:Q_series}) and (\ref{eq:f_series}) into the Riccati equation (\ref
{eq:Riccati}) we easily obtain the series coefficients $f_{j}$ as a function
of $E$ and the known potential parameters $Q_{j}$.

We rewrite the partial sums of the expansion (\ref{eq:f_series}) as rational
approximations $x^{\beta -1}[N+d/N](z)$, where $z=x^{\beta }$, in such a way
that
\begin{equation}
\lbrack N+d/N](z)=\frac{\sum_{j=0}^{N+d}a_{j}z^{j}}{\sum_{j=0}^{N}b_{j}z^{j}}%
=\sum_{j=0}^{2N+d+1}f_{j}z^{j}+\mathcal{O}(z^{2N+d+2})  \label{eq:Pade}
\end{equation}
In order to satisfy this condition the Hankel determinant $H_{D}^{d}$, with
matrix elements $f_{i+j+d-1}$, $i,j=1,2,\ldots ,D$, vanishes. Here, $%
D=N+1=2,3,\ldots $ is the determinant dimension and $d=0,1,\ldots $. The
main assumption of the Riccati--Pad\'{e} method (RPM) is that there is a
sequence of roots $E^{[D,d]}$ of $H_{D}^{d}(E)=0$ for $D=2,3,\ldots $ that
converges towards a given eigenvalue of equation (\ref{eq:Schrodinger}).
Comparison of sequences with different $d$ values is useful to estimate the
accuracy of the converged results.

Notice that we do not have to take the boundary conditions explicitly into
account in order to apply the RPM; the approach selects them automatically.
In addition to the answers expected from physical considerations, the RPM
also yields unwanted solutions as shown below.

\section{Model with bound states \label{sec:boundstates}}

From the ansatz $\varphi (r,\lambda )=r\exp \left( -r-\lambda r^{2}\right) $
and the equation $\varphi ^{\prime \prime }(r,\lambda )/\left[ 2\varphi
(r,\lambda )\right] =V(r,\lambda )-E(\lambda )$ we derive a
potential--energy function $V(r,\lambda )=-1/r+2\lambda r+2\lambda ^{2}r^{2}$
if $E(\lambda )=-1/2+3\lambda $. For $\lambda >0$ $\varphi (r,\lambda )$ and
$E(\lambda )$ are a pair of eigenfunction and eigenvalue of the Schr\"{o}%
dinger equation with the potential $V(r,\lambda )$. For $\lambda <0$ $%
E(\lambda )=-1/2+3\lambda $ is not an eigenvalue of the Schr\"{o}dinger
equation because the corresponding eigenfunction $\varphi (r,\lambda )$ is
not square integrable. Curiously enough, $e(\lambda )=-1/2-3\lambda $ is
close to the ground--state eigenvalue of the Schr\"{o}dinger equation
\begin{equation}
\psi ^{\prime \prime }(r)+2\left[ E-V_{1}(r)\right] \psi (r)=0,\;V_{1}(r)=-%
\frac{1}{r}-2\lambda r+2\lambda ^{2}r^{2},\;\lambda >0  \label{eq:V1(r)}
\end{equation}
when $\lambda $ is sufficiently small. Killingbeck \cite{K07b} calculated
the energy shift $\Delta (\lambda )=E(\lambda )-e(\lambda )$ very accurately
for several values of $\lambda $ by means of the Hill--series method.

Our interest in this model stems from the fact that $1/r-\varphi ^{\prime
}(r,-\lambda )/$ $\varphi (r,-\lambda )=1-2\lambda r$ is an exact rational
function and therefore $e(\lambda )$ will always be a root of the Hankel
determinants even though it does not correspond to a square--integrable
eigenfunction if $\lambda >0$. This unwanted solution will appear as an
exact multiple root of the Hankel determinant, very close to the physical
one when $\lambda $ is close to zero.

If $\lambda <2/27$ the potential--energy function (\ref{eq:V1(r)}) exhibits
three stationary points: a minimum at $r_{1}<0$, a maximum at $r_{2}\geq 4$
and a shallow minimum at $r_{3}>4$. On the other hand, there is only a
minimum at $r_{1}<0$ when $\lambda >2/27$. Obviously, only the stationary
points at $r>0$ make sense from a physical point of view, and we expect the
RPM to yield better results in the latter case. The expansion of the
solution to the Riccati equation about $r=0$ will require many terms in
order to take into account the shallow minimum that will move away from
origin as $\lambda $ decreases. In this case we expect to face the necessity
of Hankel determinants of greater dimension in order to obtain the shift to
a given accuracy as $\lambda $ decreases. This unfavourable situation is an
interesting test for the RPM that has not been applied to this kind of
problems before.

The Hankel determinants are polynomial functions of $\Delta $ and $\lambda $%
. For example, $H_{D}^{0}(\Delta ,\lambda )=\Delta ^{D-1}P_{D}(\Delta
,\lambda )$ and, therefore, the approximation to the energy shift is given
by a root of $P_{D}(\Delta ,\lambda )=0$ that approaches the multiple root $%
\Delta =0$ as $\lambda $ decreases. Table~\ref{tab:V1} shows $\Delta
(\lambda )$ for some values of $\lambda $ calculated with determinants of
dimension $D\leq 20$. In order to estimate the last stable digit we compared
the sequences of roots with $d=0$ and $d=1$. As expected from the argument
above, the accuracy decreases as $\lambda $ decreases if we do not increase
the maximum value of $D$ consistently, but in all cases we have verified
that there is a sequence of roots of the Hankel determinants that converge
towards the ground--state eigenvalue. Present results agree with those
calculated by Killingbeck by means of the Hill--series method\cite{K07b}.

\section{Model with no bound states \label{sec:resonances}}

It has already been shown that the RPM is a most efficient tool for the
calculation of resonances in the continuum of simple quantum--mechanical
models\cite{F95,F96b,F97}. However, for completeness in what follows we
consider the potential--energy function
\begin{equation}
V_{2}(r)=-\frac{1}{r}+2\lambda r-2\lambda ^{2}r^{2}  \label{eq:V2(r)}
\end{equation}
that is closely related to the preceding one but does not support bound
states because it is unbounded from below as $r\rightarrow \infty $. In this
case we expect unstable or resonant states with complex eigenvalues that
correspond to tunnelling from the Coulomb well.

Table~\ref{tab:V2} shows present results obtained from Hankel sequences with
$D\leq 20$. As in the preceding example we compared sequences with $d=0$ and
$d=1$ in order to estimate the last stable digit. Our results agree with
those reported by Killingbeck\cite{K07b}, except for $\lambda =0.08$. While
the first digits of the imaginary part of our eigenvalue agree with those in
Killingbeck's Table 3\cite{K07b}, the real part is completely different. The
disagreement is due to a misprint in Killingbeck's Table 3 for that
particular entry. In fact, we have found that the real part of the
eigenvalue reported by Killingbeck for $\lambda =0.08$ corresponds to $%
\lambda =0.05$ instead, as shown in present Table~\ref{tab:V2}.

\section{Conclusions \label{sec:conclusions}}

We have shown that the RPM is suitable for the accurate calculation of bound
states and resonances of perturbed Coulomb problems. The first model,
equation (\ref{eq:V1(r)}), considered in this paper exhibits interesting
features that were not faced in previous applications of the RMP\cite
{FMT89a,FMT89b,F92,FG93,F95,F95b,F95c,F96,F96b,F97}. A shallow minimum that
moves forward from origin as the potential parameter $\lambda $ decreases
makes it necessary to resort to Hankel determinants of increasing dimension
in order to obtain eigenvalues of a given accuracy. On the other hand, we
had already proved that the RPM is suitable for the calculation of
resonances in the continuum, and we simply confirmed this strength of the
approach by means of the second model chosen above.

The applicability of the RPM is not restricted to eigenvalue equations. We
have recently applied it to several ordinary nonlinear differential equations%
\cite{AF07}. Since most of them are not Riccati equations we called this
variant of the method Pad\'{e}--Hankel, but the strategy is basically the
same outlined in this paper.

\begin{table}[H]
\caption{Energy shift $\Delta(\lambda)$ for the ground--state energy of the
perturbed Coulomb model (\ref{eq:V1(r)})}
\label{tab:V1}
\begin{center}
\begin{tabular}{ll}
\hline
\multicolumn{1}{c}{$\lambda$} & \multicolumn{1}{c}{$\Delta(\lambda)$} \\
\hline
0.10 & $3.41730960373299\ 10^{-2} $ \\
0.09 & $2.31341988422733\ 10^{-2} $ \\
0.08 & $1.4212168993068\ 10{-2} $ \\
0.07 & $7.546639486534\ 10^{-3} $ \\
0.06 & $3.1738752354\ 10^{-3} $ \\
0.05 & $8.93101948\ 10^{-4} $ \\
0.04 & $1.1819718\ 10^{-4} $%
\end{tabular}
\par
\end{center}
\end{table}

\begin{table}[H]
\caption{Resonance for the $1s$ state of the perturbed Coulomb model (\ref
{eq:V2(r)})}
\label{tab:V2}
\begin{center}
\begin{tabular}{lll}
\hline
\multicolumn{1}{c}{$\lambda$} & \multicolumn{1}{c}{$\mathrm{Re}E$} &
\multicolumn{1}{c}{$\mathrm{Im}E$} \\ \hline
0.10 & $-0.27519233330828482428$ & $1.3918964850900\ 10^{-8}$ \\
0.09 & $-0.29265795893536614770$ & $7.9213310722\ 10^{-10}$ \\
0.08 & $-0.31105186469292522577$ & $2.094858859\ 10^{-11}$ \\
0.05 & $-0.372260539194895485$ & \multicolumn{1}{c}{$---$}
\end{tabular}
\par
\end{center}
\end{table}


\begin{thebibliography}{99}
\bibitem{KGJ05}  J. P. Killingbeck, A. Grosjean, and G. Jolicard, J. Phys. A
38 (2005) L695-L699.

\bibitem{KGJ06}  J. P. Killingbeck, A. Grosjean, and G. Jolicard, J. Phys. A
39 (2006) L547-L550.

\bibitem{KGJ04}  J. P. Killingbeck, A. Grosjean, and G. Jolicard, J. Phys. A
37 (2007)

\bibitem{K07}  J. P. Killinbeck, J. Phys. A 40 (2007) 9017-9024.

\bibitem{K07b}  J. P. Killingbeck, J. Phys. A 40 (2007) 2819-2824.

\bibitem{FMT89b}  F. M. Fern\'{a}ndez, Q. Ma, and R. H. Tipping, Phys. Rev.
A 40 (1989) 6149-6153.

\bibitem{FMT89a}  F. M. Fern\'{a}ndez, Q. Ma, and R. H. Tipping, Phys. Rev.
A 39 (1989) 1605-1609.

\bibitem{F92}  F. M. Fern\'{a}ndez, Phys. Lett. A 166 (1992) 173-176.

\bibitem{FG93}  F. M. Fern\'{a}ndez and R. Guardiola, J. Phys. A 26 (1993)
7169-7180.

\bibitem{F95c}  F. M. Fern\'{a}ndez, Phys. Lett. A 203 (1995) 275-278.

\bibitem{F95}  F. M. Fern\'{a}ndez, J. Phys. A 28 (1995) 4043-4051.

\bibitem{F95b}  F. M. Fern\'{a}ndez, J. Chem. Phys. 103 (1995) 6581-6585.

\bibitem{F96}  F. M. Fern\'{a}ndez, J. Phys. A 29 (1996) 3167-3177.

\bibitem{F96b}  F. M. Fern\'{a}ndez, Phys. Rev. A 54 (1996) 1206-1209.

\bibitem{F97}  F. M. Fern\'{a}ndez, Chem. Phys. Lett 281 (1997) 337-342.

\bibitem{CHS03}  H. Ciftci, R. L. Hall, and N. Saad, J. Phys. A 36 (2003)
11807-11816.

\bibitem{F04}  F. M. Fern\'{a}ndez, J. Phys. A 37 (2004) 6173-6180.

\bibitem{B05b}  T. Barakat, K. Abodayeh, and A. Mukheimer, J. Phys. A 38
(2005) 1299-1304.

\bibitem{B05}  T. Barakat, Phys. Lett. A 344 (2005) 411-417.

\bibitem{CHS05b}  H. Ciftci, R. L. Hall, and N. Saad, Phys. Lett. A 340
(2005) 388-396.

\bibitem{CHS05}  H. Ciftci, R. L. Hall, and N. Saad, J. Phys. A 38 (2005)
1147-1155.

\bibitem{F05}  F. M. Fern\'{a}ndez, Phys. Lett. A 346 (2005) 381-383.

\bibitem{AF07}  P. Amore and F. M. Fern\'{a}ndez, Rational Approximation for
Two-Point Boundary value problems, arXiv:0705.3862
\end{thebibliography}
\end{document}